# New Memory Concept for Superconducting Electronics

R. Held, J. Xu, A. Schmehl, C.W. Schneider, J. Mannhart

Experimentalphysik VI, Elektronische Korrelationen und Magnetismus,

Institut für Physik, Universität Augsburg, Germany

M.R. Beasley

Geballe Laboratory for Advanced Materials,

Stanford University, Stanford, CA 94305, USA

## Abstract

A new concept for nonvolatile superconducting memories is proposed. The devices combine ferromagnetic dots for the storage of the data and Josephson junctions for their readout. Good scalability is expected for large scale integration. First memory cells have been implemented using 3 µm-Nb-technology and permalloy dots. Nonvolatile data storage at 300 K was demonstrated.

Digital superconducting electronics (see for example Ref. 1-3) takes advantage of the fast switching time and low power consumption of Josephson junctions. Indeed, switching times in the psec-range, cycle times of 100 psec, and power dissipations of ≈ 120 nW per memory cell have been obtained in 4-kbit RAM blocks based on the RSFQ concept[4]. Digital superconducting systems have, however, not yet had major impact. This is due in part to the lack of a dense superconducting memory technology. The largest superconducting memory chips reported are only 16 kbits [4]. The integration density is limited by fundamental reasons. In current devices, the bits are stored in SQUID-loops as magnetic flux quanta. Operation of the memory cells requires the product of the critical current of the SQUID and the SQUID inductance to be comparable to a flux quantum: $I_c L \approx \Phi_0 / 2$ (Eq. 1). Since $I_c$ is limited by the need for stability against thermally induced transition between flux quanta states, Eq. 1 requires that $L > 10$ pH. For a typical SQUID design, this inductance corresponds to a loop diameter of several micrometers, which leads concomitantly to the large size of the memory cells. To avoid this problem for superconducting electronics, hybrid Josephson-CMOS memories are now being developed, in which the data are stored in semiconducting CMOS cells, and are read out by high speed superconducting detectors[5, 6]. Here we propose a new way to overcome this fundamental size limitation of superconducting memories that results from the size of the flux quantum. In this approach, superconductors are combined with ferromagnetic materials that have high permanent moments in order to shrink the area required by the stored flux. In a sense, this approach is the magnetic analog of using CMOS with Josephson junction readout. Here we are using MRAM with a Josephson junction readout.

This concept is illustrated in Fig. 1 where we show a possible implementation of such a memory element. To write the data into the cells, a magnetic field is generated by a current pulse flowing in a superconducting write-line that magnetizes a ferromagnetic dot located adjacent to a Josephson junction. After writing, the ferromagnetic dot is magnetized and its field $H_{FM}$ penetrates the junction barrier. Depending on the direction of the magnetizing current pulse, $H_{FM}$ points either away from the junction, corresponding to, say, a logical "0", or towards the junction, corresponding to a "1". To read the magnetization direction, a magnetic background field $H_{BG}$ is generated by a dc current flowing under the Josephson junction through a superconducting background-line. Since the absolute value of the magnetic field in the junction $H = |H_{BG} + H_{FM}|$ depends on the direction of $H_{FM}$, the polarity of the magnetization can be derived from the magnetic-field-dependent critical current $I_c$ of the junction.

What are the requirements for the sizes of the junction and the dot? For a straightforward estimation we approximate the magnetic field of a square ferromagnetic dot of side-length $l$ and thickness $d$ by a dipole field of a corresponding single domain particle with magnetization $m$. As a function of the distance $a$ between the dot and the junction, the dipolar field of the dot varies as $m_r V/a^3$, $m_r$ being the remanent magnetization and $V$ the volume of the dot. In practical implementation $a \approx l$, so that the magnetic flux penetrating the junction barrier is $\Phi \cong \mu_0 m_r d \lambda_L/\pi$. Here $\lambda_L$ is the superconducting magnetic penetration depth, which equals 90 nm for Nb films at 4.2 K. Thus, magnetic particles with remanent magnetizations of 0.18 T will generate a flux $\Phi \approx \Phi_0/2$ in the junction barrier, the desired scale for device operation. It is a key advantage of our new memory concept that the flux $\Phi$ seen by the Josephson junction does not depend on $l$ and that therefore the basic principle of operation does not impose a lower limit on the size of the cells.

Will magnetic crosstalk between neighboring cells be a problem? The magnetic field generated in the Josephson junction of cell 1 by the ferromagnetic dot of cell 2 scales as $(1/L)^3$, where $L$ is the distance between the dot of cell 1 and the junction of cell 2. For a design with $L \approx 4\, l$, crosstalk is expected to be unimportant. We therefore conclude that the lower limit of the device size is likely determined by the minimum size of either the Josephson junction or of the magnetic dot, which suggests the possibility of very dense memories.

To test the concept, first devices were designed and fabricated. These use Josephson junction arrays, each containing eight junctions. The arrays were fabricated by Hypres, using a 3 µm, 1 kA/cm$^2$ Nb-technology[7]. The write-lines and the background-lines, which were 10 µm and 5 µm wide, respectively, were also made from Nb (problems with the write-lines caused us to shunt the Nb of the lines with Au cap layers). To keep the design simple, for each array only one write-line was used and all bits of an array were simultaneously selected.

For the ferromagnet a wide selection of materials is available. For the first devices permalloy (Py, $Ni_{81}Fe_{19}$) was chosen. The dots were deposited on the previously fabricated Nb junction array chips at room temperature by rf-sputtering Py. The ferromagnetic films were patterned by e-beam lithography and lift-off. An optical micrograph of one of our memory cells is shown in Fig. 2.

Control samples were used to measure the magnetic properties of the Py dots by SQUID-

magnetomety. A typical $m(H)$ characteristic is shown in Fig. 3. The relative small remanence (0.30 T) and large coercitivity (14 mT) of the Py dots is ascribed to the growth-induced microstructure of the Py films. The growth conditions were chosen so as not to damage the Josephson-junctions, rather than to optimize the magnetic properties of the Py.

The current-voltage characteristics ($I(V)$) and the $I_c(H)$ dependencies of the junctions followed the specifications and were standard for Nb tunnel-junctions. They change neither as a result of the sputtering nor of the patterning of the Py. The low-field $I_c(H)$-characteristic of a junction without Py dot is shown by curve A in Fig. 4. This characteristic changed if a Py dot was added and magnetized by write current pulses. Curves B and C show the corresponding characteristics for the two current pulse directions. Due to the dot´s magnetic field, the center of the curves is shifted from $H = 0$. The maximum $I_c$ is suppressed, presumably due to field inhomogeneities and vortices trapped in the Nb-layers. To investigate whether the $I_c(H)$ curves were indeed being controlled by the magnetization of the permalloy and not merely by vortices generated by the write pulse and trapped by the Nb, the chip was cycled to 300 K. Since the $I_c(H)$ curves did not change after the cycling, we conclude that the shifts are caused by the remanence of the Py dot.

With the help of a small field generated by the background-line, memory function was achieved. This is demonstrated in Fig. 5, in which we show the junction's critical current as modified by the preceding write pulses. The time scale of these measurements is determined by the acquisition time of the electronics used. Since the time scales for the magnetization of small ferromagnetic dots and for the switching of Josephson junctions are very fast, we expect that optimized devices operate at high speeds.

As implemented, the write currents needed to switch the permalloy were as high as 100 mA, much larger than desired. Since the required currents depend on the coercitivity of the ferromagnet and on the architecture of the device, we are confident that they can be considerably reduced. We furthermore noted that trapping of magnetic flux lines, predominantly caused by the large write pulses, tended to affect the junction characteristic. We expect that also this flux trapping can be reduced by optimizing the cell design.

In summary, a new concept for a superconducting memory has been suggested and successfully demonstrated. The memory cells use ferromagnetic dots for storage and Josephson junctions for readout. We expect that the devices can be integrated on a large scale and operated at high speed. They have the further advantage of being nonvolatile.

The authors thank A. Herrnberger for his support. J.M. thanks the Geballe Laboratory for Advanced Materials and the colleagues their for their hospitality to enable a visit during which the ideas underlying this work were developed. The authors furthermore acknowledge support by the BMBF / VDI project 13N6918, and the US Air Force Office of Scientific Research.

FIGURE CAPTIONS

**Fig. 1:** Schematic drawing of the operation of a single memory cell. The magnetic field of a superconducting write-line magnetizes a ferromagnetic dot. The remanent magnetization of the dot controls the critical current of a Josephson tunnel junction.

**Fig. 2:** Optical micrograph of first memory devices with closeup view of a single memory cell (top). The ferromagnetic dot has the dimensions 6 µm x 9 µm x 600 nm.

**Fig. 3:** Typical ferromagnetic hysteresis loop ($m(H)$ characteristic) of an 18x18 dot array measured at 5 K. The dots were 6 µm x 9 µm x 600 nm in size.

**Fig. 4:** $I_c(H)$ characteristic of a Josephson junction without permalloy dot (curve A) and with magnetized permalloy dot (curves B and C) measured at 4.2 K. The write-current pulses preceding the measurements of curves B and C, respectively, had opposite polarity. The sample was cycled to 300 K between the write and read procedures.

**Fig. 5:** Demonstration of the memory function of a cell. The upper trace shows the write-currents, the lower curve the measured critical current of the Josephson junction (4.2 K).

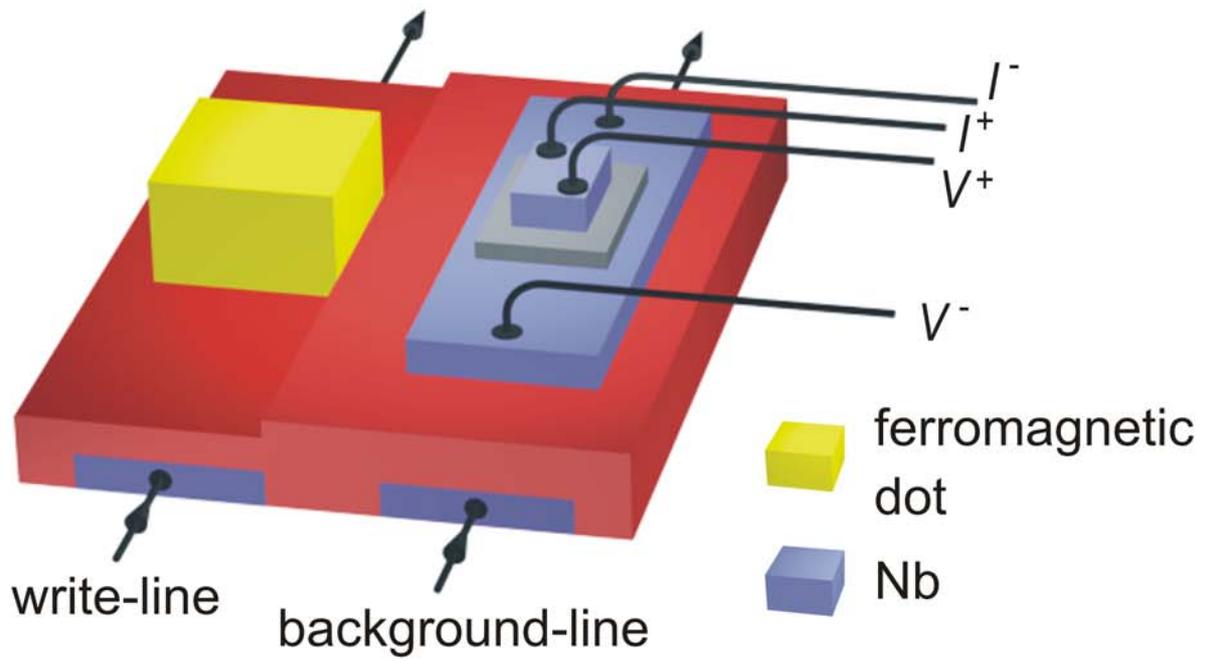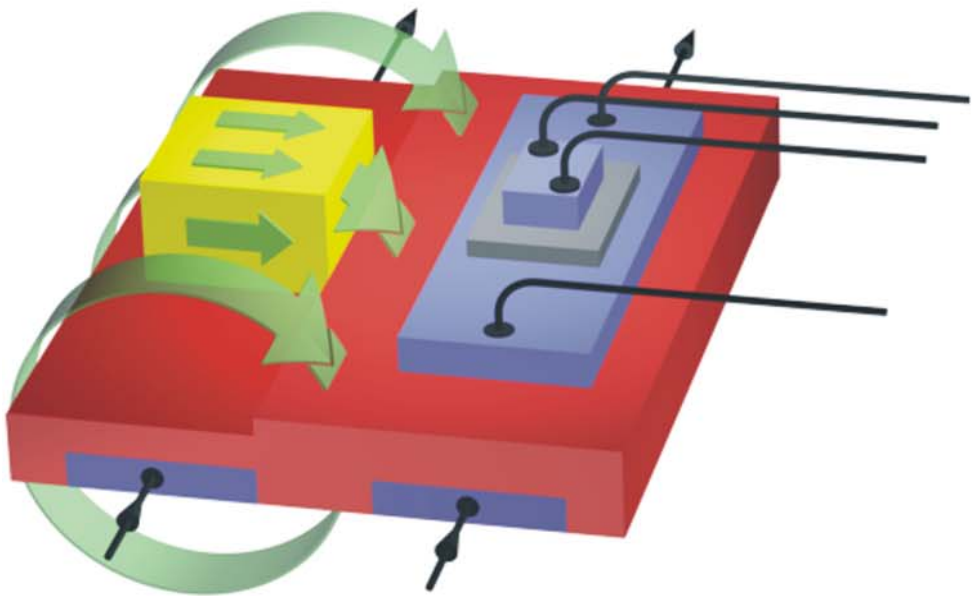

Held et al., Fig. 1

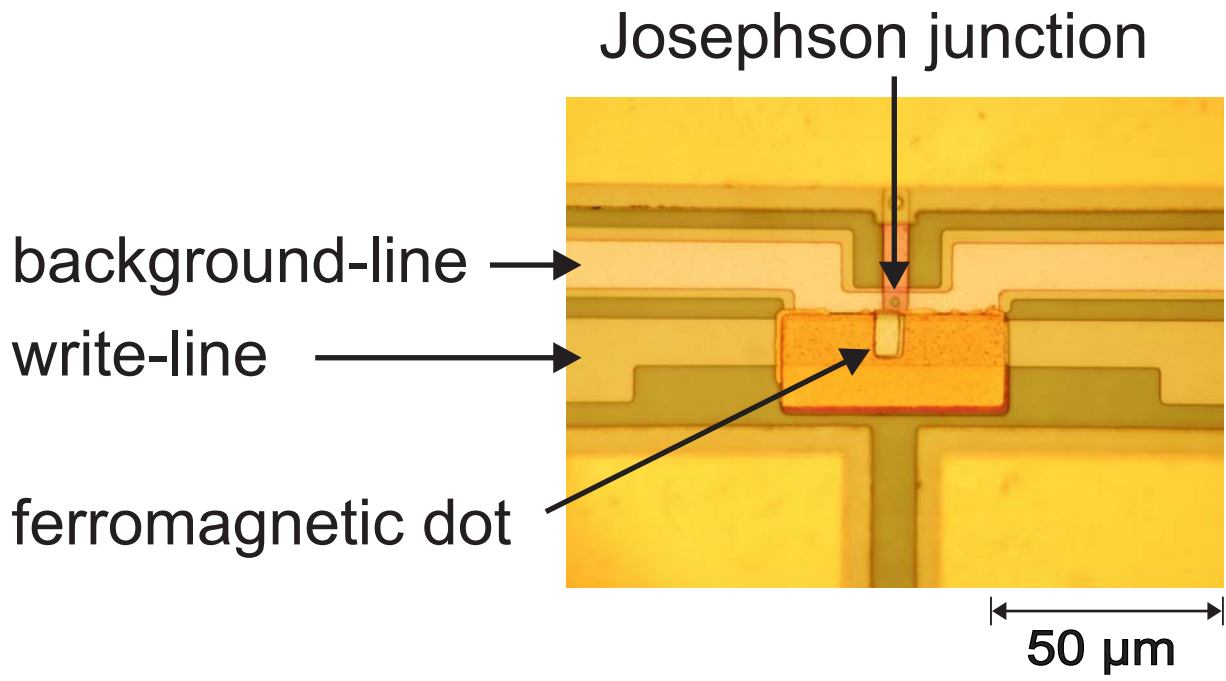

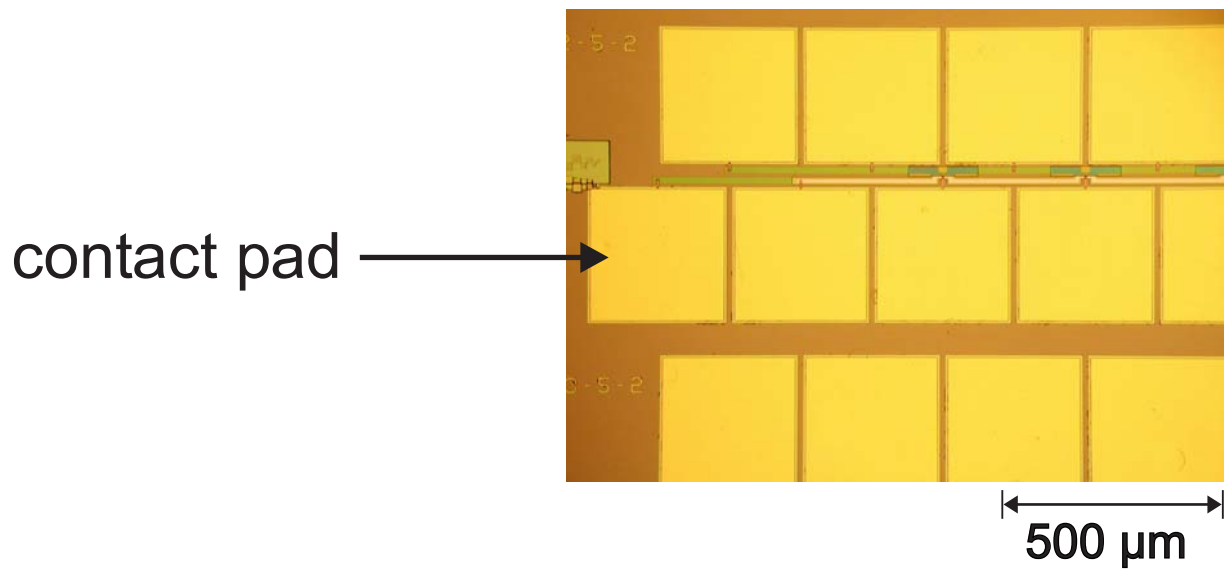

Held et al., Fig. 2

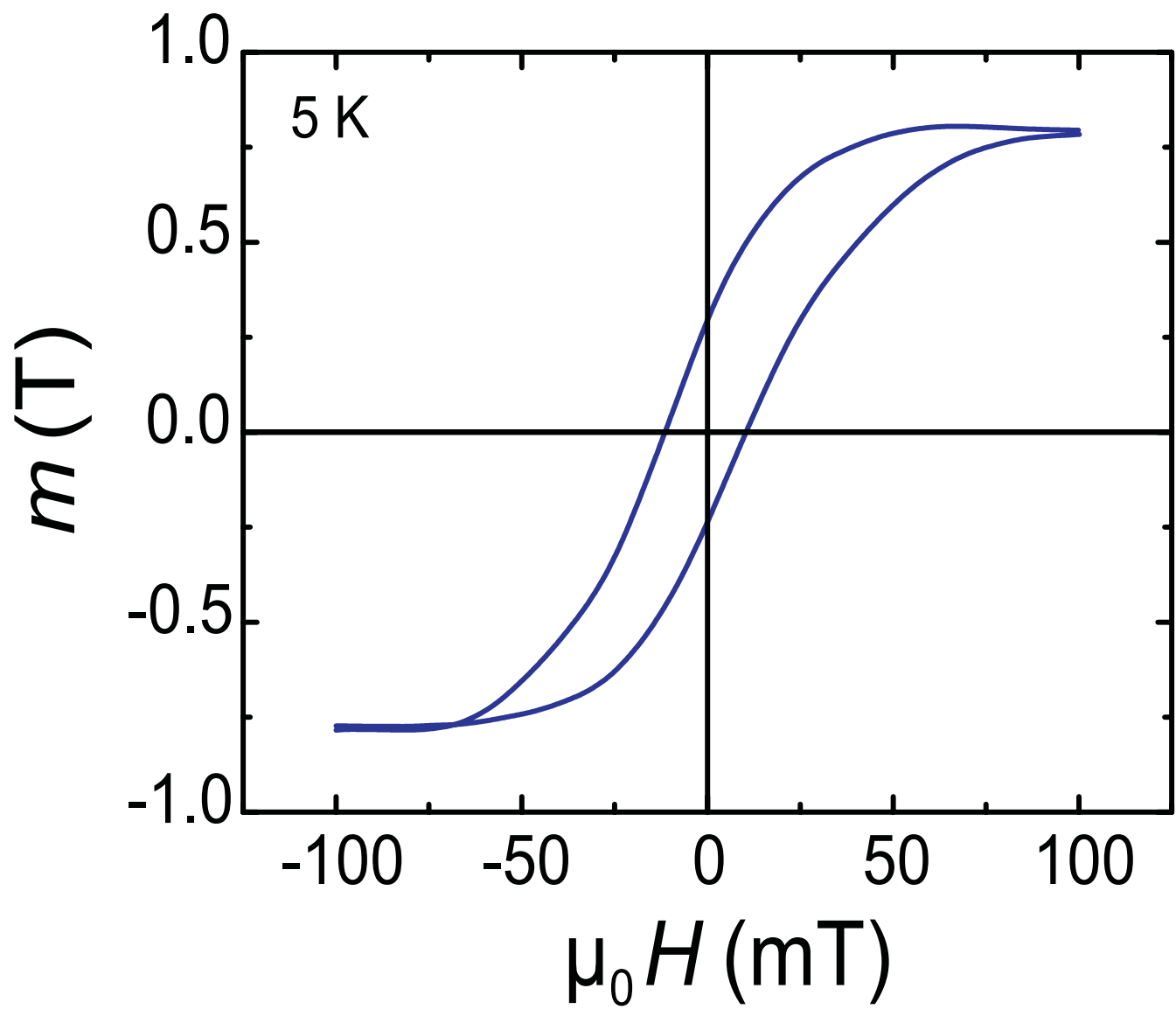

Held et al., Fig. 3

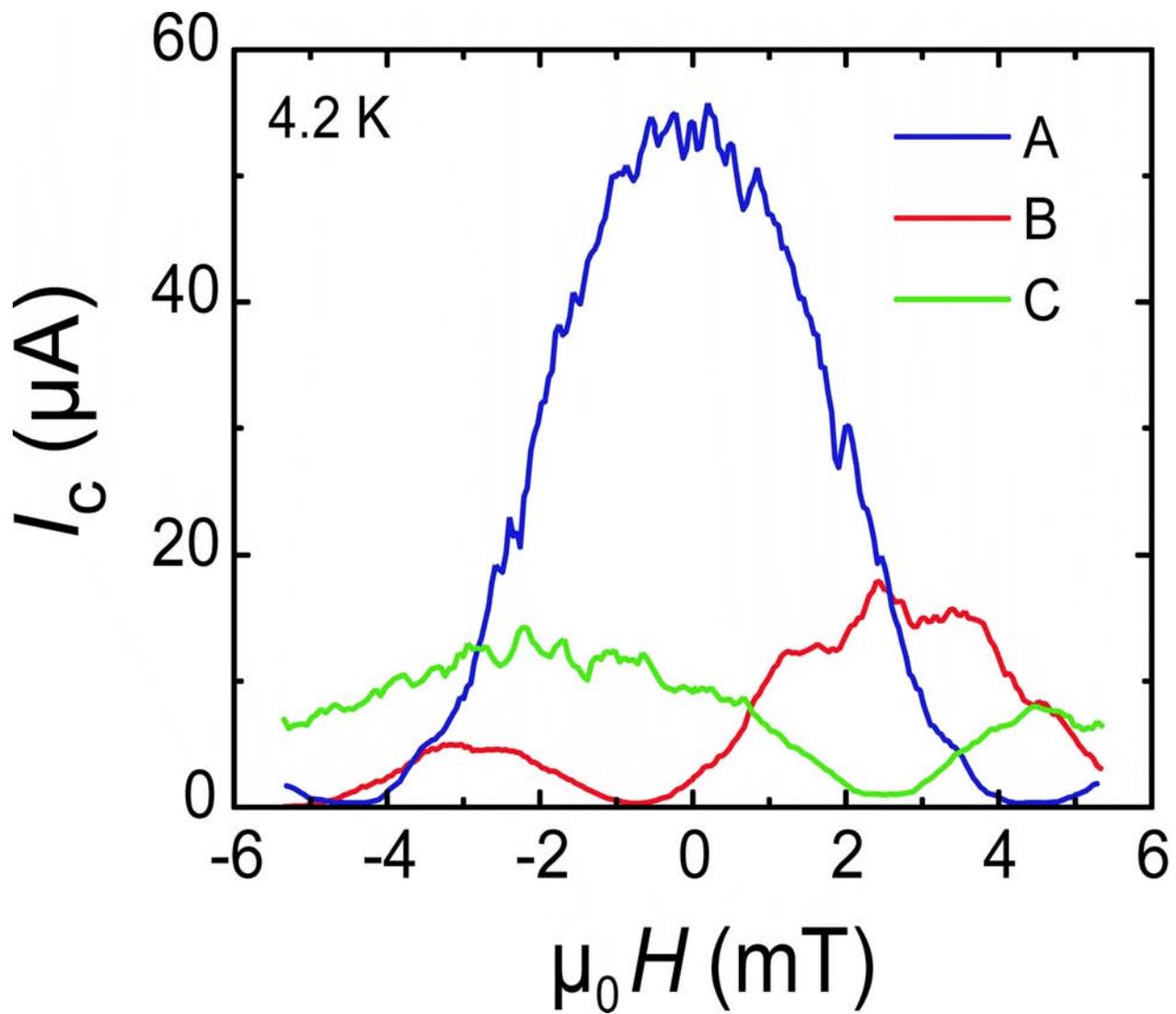

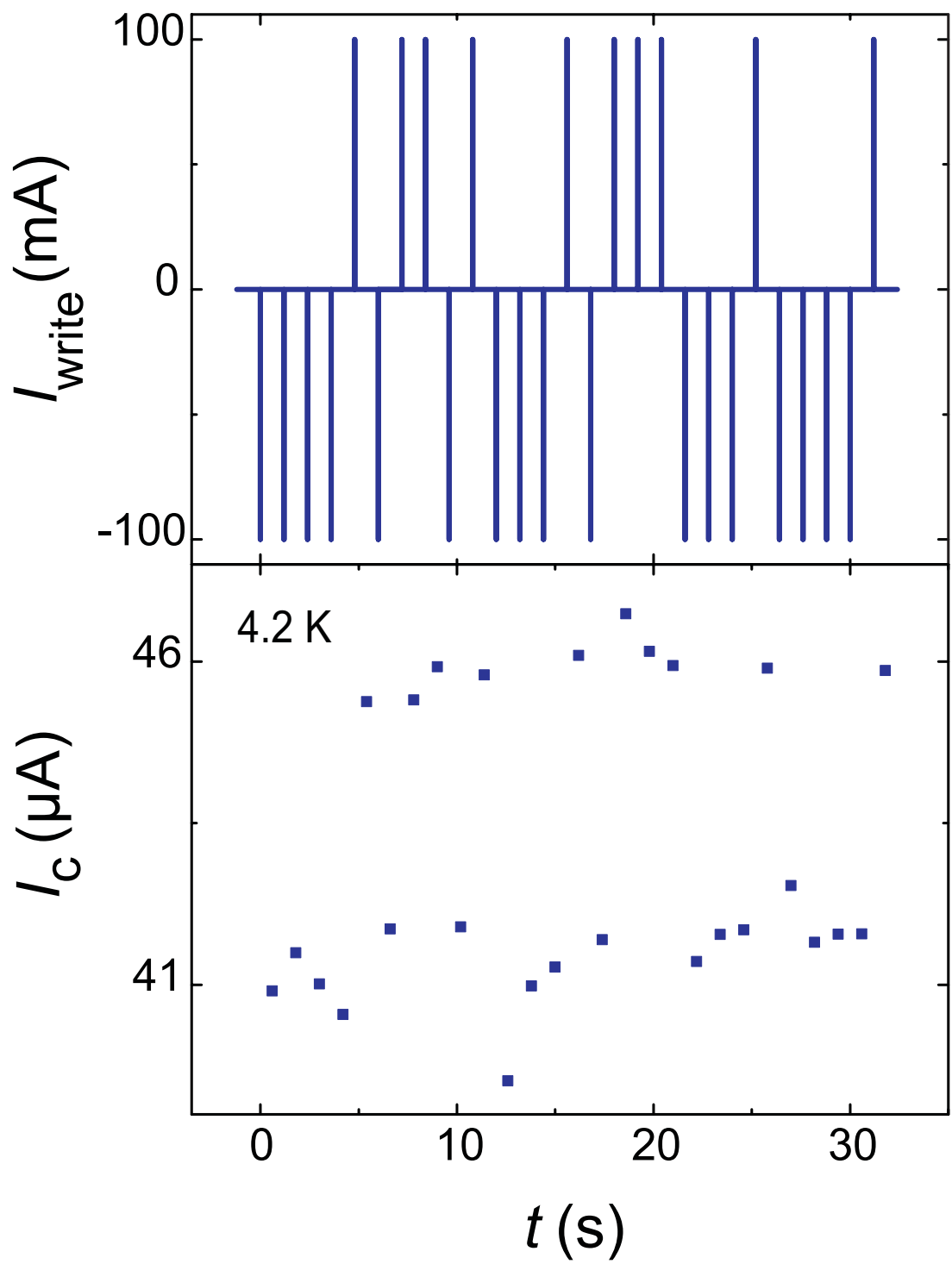

Held et al., Fig. 5